Are BL Lacertae Objects Beamed QSO Remnants?

E.F. Borra

Centre d'Optique, Photonique et Laser, Observatoire Astronomique du Mont Mégantic,
Département de Physique, Université Laval, Québec, Canada G1K 7P4

**ABSTRACT**

This paper considers the hypothesis that BL Lacertae objects (BLLs) are the beamed remnants of Quasi Stellar Objects.  The hypothesis explains why BLLs do not undergo the strong evolution seen in other active galactic nuclei since it naturally  predicts that the space density of BLLs should *increase*  with cosmic time, as shown by recent observations. Numerical models reproduce, with reasonable parameters, the known redshift and magnitude counts of BL Lac objects. It is assumed that radio-quiet  as well as radio-loud quasars are capable of generating jets but that jets are snuffed in young radio-quiet objects and only emerge in aged ones. I argue that the observations allow this assumption.

# 1. INTRODUCTION

It is increasingly accepted that the objects that populate the high-energy extragalactic "zoo" are Active Galactic Nuclei (AGNs), that they are manifestations of the same basic phenomenon and that they all will eventually be unified within energetic, geometrical or evolutionary schemes. Among AGNs, the BL Lacertae objects (hereafter referred to as BLLs) are particularly puzzling "beasts", since they exhibit extreme characteristics among the AGNs. A consensus has emerged that these characteristics are due to relativistic beaming[1]. Over the past several years, evidence has gradually built up showing a puzzling behavior that perhaps differentiates them the most from other AGNs: They do not show the large decrease in comoving density with cosmic time exhibited by other AGNs. On the contrary, Morris et al.[2] recently found evidence of negative evolution, namely that the space density of BLLs actually *increases* with cosmic time.

Borra [3] has proposed that BL Lacertae objects are the beamed remnants of Quasi Stellar Objects (QSOs).This paper was probably not noticed at that time by workers active in AGN research because the hypothesis was used in a cosmological test rather than presented as the theme of the paper. What prompted this hypothesis was that it naturally explains why BLLs do not undergo the strong evolution seen in other AGNs; indeed negative evolution is a natural consequence of it. Borra [3] made the assumption that all Quasi Stellar Objects (QSOs), radio-quiet (RQQ) as well as radio-loud (RLQ), are capable of generating jets but that jets are snuffed in young RQQs and only emerge in aged objects. This assumption was necessary to obtain good agreement between the space densities of the objects concerned. Prima facie, this assumption runs counter the prevailing belief that radio-quiet and radio-loud quasars are distinct objects hosted in different types of galaxies (spirals versus ellipticals) located in different environments (poor versus rich clusters of galaxies). On the other hand, radio-loud and radio-quiet QSOs have remarkably similar spectra and powers, so that it is natural to assume that the same basic mechanisms are at work and that, for some reason, energy in the radio region is either absorbed or not generated in the RQQs. Section III further addresses this issue.

In this paper, I reexamine the hypothesis in the light of 10 years of advances in our knowledge of AGNs. I assume that all QSOs have jets but that the jet is quenched in the young RQQs. As the QSO ages, the quenching mechanism weakens allowing the jet to emerge. I shall also argue that the observational evidence on the environmental statistics of the RLQs and RQQs does not rule out the hypothesis.

## 2. NUMERICAL MODELS

The basic hypothesis assumes snuffing that decreases in strength with age but does not identify a specific snuffing model. Let us simply assume that all AGNs are capable of generating a jet and that material surrounding the nucleus is somehow responsible for snuffing it in radio-quiet QSOs. This material may be in a leaky shell, or torus, surrounding the central powerhouse, with holes or weak spots that allow the emergence of a jet only if it is aimed at one of them. With a toroidal geometry, the jet can be directed along directions not necessarily aligned with the axis of symmetry of the torus. We shall not make any assumption on the snuffing mechanism itself but assume that the material interferes somehow with the mechanism that either accelerates the material or delivers it to the outer extended lobes. It seems reasonable to assume that the quantity of material surrounding the nucleus eventually decreases as the QSO ages, so that it becomes easier for a jet to emerge from an older quasar.

One could in principle evolve the observed luminosity function of QSOs from high to low redshifts, compute the rate of formation of remnants and compare the predictions to the observed luminosity function of the local BL Lac objects. However, in the absence of a quantitative theory, this is not a particularly fruitful approach, and we shall instead simply find a local luminosity of the QRs that is compatible with the known evolution of the luminosity function of QSOs and the local luminosity function of BLLs. We then shall see whether relativistic beaming can give a luminosity function that reproduces the observations.

Urry & Shafer [4] have studied the effects of relativistic beaming on a luminosity function. Following [4], we shall assume a power law for the isotropic unbeamed component of the luminosity function of the local QSO remnants (QR):

$$\Phi_{QR}(L,z=0) \, dL = KL^{-B} \, dL \quad \text{for} \quad L_1 < L < L_2$$
$$= 0 \quad \text{for} \quad L < L_1 \text{ or } L < L_2 \quad . \quad (1)$$

As Urry & Shafer [4] point out, the upper and lower luminosity cutoffs, although unrealistic, approximate the fact the luminosity function must turn down at some luminosity. They also show that relativistic beaming changes the original power law luminosity function in a way that can be approximated by two power laws, the high luminosity end having index ~ B, and the low luminosity end having a flattened exponent ~ (p+1)/p, where p depends on the shape of the spectrum, the structure of the jet and the frequencies being compared; it usually is in the range 3<p<5. We keep the notation of [4] and the reader is referred to that paper for more details.

If QSOs are long lived, the local density of their remnants should be essentially equal to the density of quasars at z = 2 and

$$\int_{M_{lr}}^{M_{ur}} \Phi_{QR}(M,z=0.0) dM = \int_{M_{lq}}^{M_{uq}} \Phi_{QSO}(M,z=2.0) dM \quad , \quad (2)$$

where $\Phi_{QR}(M,z)$ and $\Phi_{QSO}(M,z)$ are the luminosity functions of QSO remnants and QSOs. This approximates the actual situation since QSOs are probably continuously formed rather than in a burst at z= 2.0 as implied by Equation 2. The $M_u$ and $M_l$ upper and lower magnitude cut-offs take into account the fact that the luminosity functions must turn down at some magnitudes and that very faint QSOs probably do not evolve significantly to our epoch. The unbeamed luminosity function is further restricted by the requirement that local QSO remnants are inconspicuous and therefore much less luminous than QSOs. Therefore we must have, at least for $M_V < -22$

$$\Phi_{QR}(M,z=0) << \Phi_{QSO}(M,z=0). \qquad (3)$$

We now feel free to experiment with luminosity functions of the form given by Equation 7 in [4] subject only to the conditions set by Equations 2 and 3.

The redshift distribution given in [2] has two peculiarities: It shows a very sharp drop at $z < 0.2$ and a more gradual, but still rapid, decrease for $z > 0.2$. A narrow luminosity function having the appropriate mean luminosity, combined with negative evolution of the space density of the objects can reproduce this. Such a luminosity function naturally arises from a beamed population of low-luminosity galactic nuclei having a power-law luminosity function with a steep index and small luminosity range. The small luminosity range may be real and approximate a sharply peaked luminosity function (e.g., Gaussian-like). One also may justify the low-luminosity cut-off by the fact that a faint object no longer looks like a BL Lac object since there is a strong contribution to the spectrum from the host galaxy; while the high luminosity drop-off approximates the fact that the luminosity function must turn down at some luminosity. It is also important that the unbeamed objects be faint since too high a mean luminosity would shift the peak of the redshift distribution to high redshifts.

We obtain a good fit to the observations with a model having an unbeamed power-law luminosity function with index B= 3.0 and $-11 < M_V > -13$. This luminosity function respects the requirements set by Equation 3, that the unbeamed component be inconspicuous. We obtain the normalization factor K from Equation 2, where we use the luminosity function $\Phi_{QSO}(M,z)$ determined in [5] for $q_0$=0.5 and $H_0$=50 Km/sec/Mpc. We use $M_{lq}$=-22 and $M_{uuq}$=-30, thus assuming that QSOs fainter than $M_B$=-22 do not evolve enough to z =0 to contribute significantly to the local population of QRs. Although the luminosity functions in [5] are truncated at $M_B$=-23 they show strong evolution at $M_B$ = -23, and are compatible with significant evolution at $M_B$=-22. There are sources of uncertainty in K, arising from the determination of the luminosity function of QSOs, from uncertainties in the cosmological parameters and from the lower magnitude cutoff. We use a relativistic Lorentz factor $\gamma = 10$, which is a reasonable average value [6], and p= 4 for the exponent in Equation 1, a value appropriate for our problem. The beamed luminosity function is computed assuming viewing angles $0.0° < \theta < 5°$. The beamed luminosity function has a steep power law from $-26 < M_V < -24$ joined to a shallow one for $-24 < M_V < -21.5$, compatible with the known luminosities of BL Lac objects. The $M_V < -21.5$ low-luminosity cutoff justifies truncating the viewing angle at 5° for the luminosity would then be less than the luminosity of the host galaxy (BLLs appear in bright host galaxies) and the object would no longer be identified as a BLL. We compute the redshift distribution from the usual cosmological relation given by

$$\frac{dN}{d\Omega dz} = \int_{m0}^{m1} \Phi_{BL}(M,z)\frac{dV}{dz}dm \quad , \qquad (4)$$

where $\Phi_{BL}(M,z)$ is the beamed differential luminosity function (per unit magnitude), $\Omega$ the surface area and $dV(H_0,q_0,z)$ the cosmological volume element. The z dependence of the luminosity function is given by

$$\Phi_{BL}(M,z) = \Phi_{BL}(M,z=0)(1+z)^{\beta} , \qquad (5)$$

following the density evolution law determined in [2]. We shall use $\beta = -5.5$, the best fit values determined by them for density evolution. The sample in [2] is X-ray selected and, in principle, it would be desirable to base our computations fully on the X-ray luminosity function of QSOs. In practice, because one needs the luminosity functions of QSOs at z = 0 and z = 2.0 (Equations 2 and 3) and because they are much better known in the optical, especially at low fluxes and high redshifts, it is preferable to use optical luminosity functions. For our computations, we shall use the $m_V$ distribution given in [2] for their sample and a value of $m_l = 20$ in Equation 4 as indicated from their $m_V$ magnitude distribution. We use $H_0 = 50$ Km/sec/Mpc and $q_0 = 0.5$. The model described gives the redshift distribution of Figure 1 (dashed line), where the histogram gives the observations in [2]. We can see that the model fits well the main features of the redshift distribution in [2] and it is reasonable to assume that the observations sample the computed parent distribution of Figure 1. The computed z-distribution rises sharply, peaks at the same redshift and rapidly drops to undetectability for z>0.8.

We have plotted in Figure 2 the integrated counts as a function of magnitude obtained from

$$\frac{dN}{d\Omega dm} = \int_{z_0}^{z_1} \Phi_{BL}(M,z)\frac{dV}{dz}dz \quad , \qquad (6)$$

where the integrand is the same as in Equation 6 and all symbols and constants are the same. We have plotted in the same figure the estimates of the optical integrated counts of BLLs obtained in [6]; the arrows indicate either lower or upper limits.  We can see that the theoretical counts also are in agreement with the observations.

Our model is a simple one since it considers the contribution of the beam alone, assumes that all sources have the same value of γ, and uses a power-law luminosity function. It does not take into account that, in the optical, there also is a contribution from the light of the host galaxy. Padovani and Urry [6] have considered this case and find that it does not change the basic conclusions of the simple model. Their modified model, that uses two additional free parameters, still predicts a double power law with slope ~B above the break but with slope steeper  than (p+1)/p below the break. They have also extended their basic model [7] to more general luminosity functions and the case where there is a distribution of Lorentz factors γ. The uncertainties in the observations and some of the assumptions that are made in the computations do not justify the introduction of additional free parameters into our models nor the consideration of other types of luminosity functions.

## 3. DO RADIO-QUIET QUASARS HAVE SNUFFED JETS?

a) Snuffing and energy distributions

RLQs and RQQs have remarkably similar spectral energy distributions and powers at all wavelengths except radio wavelengths; it is therefore logical to assume that they actually are similar objects and that the remarkable differences in their radio fluxes are due to absorption of the radio power  or snuffing of the radio emission mechanism.  Another attractive model invokes relativistic beaming and different viewing angles to explain the difference but  fails ([8, 9, 10]). A number of absorption processes have been considered (e.g., synchrotron self-absorption, free-free absorption) but fail to account for the bimodal distribution of radiopower in quasars [8]. The remaining viable hypothesis, also suggested by  [9], is that some mechanism prevents a relativistic jet from fully developing and generating the relativistic electrons responsible for the radio emission. Snuffing naturally explains the remarkable similarity of the 2 types of quasars at all wavelengths but the radio region, as well as the bimodality of the radio emission. It accounts for the observations of [11]  who find that lobe dominated radio quasars have spectra that show deep minima in the millimeter region, indicating thus that their radio emission is not a continuation of the optical-infrared spectrum but that they contain normal RQQ cores with additional emission from a morphologically unrelated radio source.

b) Changes in the number ratios (RLQ/RQQ) with z  and magnitude

Under our hypothesis radio-jets emerge as the RQQs age, leading to the prediction that the fraction of radio-loud objects should increase with age, hence decreasing z. Therefore one could, in principle, use radio observations of optically selected QSOs to verify these predictions. However, an observational test is difficult because we simply do not know how the snuffing factor varies with age; hence we cannot quantitatively predict how the ratio RLQ/RQQ varies with redshift. Furthermore, if the jets emerge when the objects are faint, they will not longer be identified as quasars, rendering the comparison meaningless. The redshift distribution of the X-ray selected BLLs gives some clue, since the counts peak at z~0.3, first indicating the cosmic time at which most radio jets emerge and second that jets emerge mostly among very evolved and very faint objects that would not be recognized as quasars because of their very low luminosities. Comparison to the observations must therefore be tempered by these realizations. On the other hand, it is legitimate to assume that, although jets emerge very late in the evolution of most RQQs, the emergence of the jets may occur earlier in some objects and there may therefore be a tendency for the fraction of RLQs to increase with cosmic time. Whether this should be detectable and is detected with the surveys presently available is another matter that we shall now consider. This preamble should therefore make it clear that our hypothesis would stand even if the test of this subsection fails.

There are very few surveys of optically selected QSOs. The most recent and largest ones are by [10] and [12]. They confirm the bimodality of the radio emission in QSOs. Visnovsky et al. [12], who also use the data in [10] find strong evidence that in optically selected QSOs the radio-loud fraction increases with decreasing redshift, in agreement with our prediction but also that the ratio decreases with fainter limiting magnitude (albeit with weaker statistical significance). Under our hypothesis, the ratio loud/quiet should increase with decreasing redshift in optically selected samples, because at a given magnitude, as the universe ages, one collects an increasing number of older, originally brighter QSOs.

Because quasars get fainter as they age, one would expect, prima facie, that the ratio should also increase with faintness; however, this prediction may be too simple-minded. If brighter sources have a higher probability of developing a jet either simply because the jet is stronger and can get through a snuffing shield more easily or because the snuffing shield is ablated faster by the stronger radiation from the brighter core, this will cause the RLQ/RQQ ratio to increase with brightness for middle-aged objects. This picture finds some support from [12] where correlation between radio and optical luminosity for the radio-weak sources is found, as though more luminous objects have a stronger embryonic radio-jet trying to break through and giving some radio flux prior to snuffing. Optically selected QSOs are found in surveys that are magnitude limited, the absolute flux decreasing with z. They therefore include only relatively bright (typically MB <-23), presumably middle-aged objects, missing thus the very aged objects that have all developed jets, because they are too faint. Visnovsky et al. [12], mostly discuss QSOs with MB<-23 and 0.1<z< 3. It is possible that in the range of brightness and redshifts discussed in [12], the brightness effect dominates over the aging effect thus explaining why the ratio RLQ/RQQ increases with brightness.

I do not claim that this discussion proves that RQQs have snuffed jet but only that the observations do not rule this hypothesis. Section 3b is particularly speculative but , on the other hand, it could be dropped without affecting our discussion. The stronger argument in the favor of the snuffing hypothesis is by absentia of an alternative.

## 4. DISCUSSION

Prima Facie, the weakness of this work comes from the assumption that RLQs and RQQs are basically the same objects, which runs counter the presently held belief that these are different objects, residing in different environments (RLQs in elliptical galaxies and rich clusters, RQQs in disk galaxies and poor clusters). However, first, this is not a firmly established fact since the statistics are poor and the earthbound observations have too marginal a resolution to unambiguously reveal the morphologies of the host galaxies. The uncertainties in the properties of the host galaxies have been studied [13], reaching the conclusion that, although they favor the hypothesis that RLQs are in elliptical hosts and RQQs in disks, the uncertainties are considerable. A second consideration comes from the fact that the meager observational data are obtained at z < 0.6 and one has no information about the morphologies of the host galaxies of RLQs and RQQs at higher redshifts. Third, recent spectroscopic and morphological observations coming from an HST key project carried out by Butcher, Dressler, Oemler and Gunn (reported in [14]) of 2 clusters at z = 0.4 find a much greater proportion of disk galaxies in those clusters than there are at low z. They also find that they often appear a bit disturbed or irregular in morphology. Dressler concludes that some mechanism, presumably interactions and mergers, is responsible for the disappearance of the spiral galaxies in clusters since z = 0.6. This is relevant to this work since it allows us to assume that disk galaxies hosting RQQs at high z have changed morphology to ellipticals at low z. In conclusion, the observational evidence therefore allows us two assumptions: First that at z ˜ 2 any Hubble type can host RQQs as well as RLQs; even though RLQs may favor ellipticals and RQQs disk galaxies at low redshifts. One may of course wonder why it should be so; one can only speculate that this is due to a younger universe and younger environments. Second, the HST observations also allow another plausible assumption; that the host spirals at high z have now become ellipticals. The  HST observations makes this assumption the most plausible one.

Until recently, it was generally believed that the hosts of BLLs were exclusively elliptical galaxies. However, there is now conclusive evidence that at least some BLLs are in disk galaxies. For example, Abraham, McHardy, & Crawford [15] have carried out an optical imaging survey of BLL host galaxies, concluding that out of 6 classifiable hosts 2 are disk systems. One may be skeptical of this classification since there have been previous controversies regarding the nature of the hosts of some BLL (e.g. [16], and references therein). However, an unpublished high-quality HST image of the BLL PKS1413+135 (McHardy, private communication) unambiguously shows it to be a in a disk galaxy. As [15] points out, the discovery of disk hosts is in disagreement with the standard model that assumes that all BLLs are hosted in elliptical galaxies . This is however allowed by our hypothesis.

There have been proposals to relate BLLs to RLQs ([2, 17]). With respect to our hypothesis, these have the advantage of not having to assume a snuffing mechanism since they relate only radio-loud objects. However, because the ratio RQQ/RLQ is  about 10 at z =2, it is more difficult to reproduce the observed space density of BLLs at low z and obtain the sort of good agreement with observations seen in Figures 1 and 2. It is of course possible to reduce the gap by assuming parent quasars having Mv >-22 , by decreasing the relativistic γ factor or by taking the lower limits to the space densities of BLLs from [2]. However, this fine tuning must be contrasted with the agreement that we obtained with more reasonable estimates. Finally, the assumption that RLQs are exclusively in elliptical galaxies and are the progenitors of BLLs is in conflict with the presence of some BLLs in disk galaxies.

There obviously are details that our basic hypothesis does not explain. For example, one may want to explain the relation between OVV quasars (radio quasars with flat radio spectra and highly polarized QSOs) and BLLs and their spectral differences (e.g. [18, 19]). This is beyond the scope of this work, but it is reasonable to assume that the spectral differences are due to different evolutionary ages or environments . Assuming that BL Lacs are more evolved objects than quasars, it is not surprising that there are spectral and morphological differences between them and similar objects they may be evolutionary related to. It is the reverse that would be surprising. In particular, our hypothesis is not in conflict with the assumption that the parent population of BL Lacs is a subset of FR I. We can assume that although strong and young RLQs are FR II, the weaker and more evolved BL Lacs are FRI. Perhaps it is power that decides whether an object is FR I or FR II.

Finally, it is interesting to notice that the leaky-shell, leaky-torus model may also explain why BLLs have very weak broad emission lines. If the shell contains enough absorbing material (e.g., dust), and if it surrounds the broad-line region, it can greatly absorb the flux from the region that forms the broad lines. Although the jet punches a hole through the shell, the hole may be large enough to let the jet through but too small to allow a significant flux from the emission-line region to give strong broad emission lines. In a classical QSO, the viewing angle would be such that the observer sees the nucleus and broad-line region through a large hole or at a small angle with respect to the axis of symmetry of the torus. Old age may explain why narrow lines are weak. While a young QSO should have a healthy narrow-line region, since it is in a young galactic environment with plenty of gas, a BLL does not have a strong narrow line region since gas is depleted in its older environment.

## 5. CONCLUSION

We have reexamined, after ten years of advances in AGN research, the hypothesis advanced by Borra [3] that BL Lacertae Objects are the beamed remnants of Quasi Stellar Objects. This hypothesis explains why BLLs do not undergo the strong evolution seen in other AGNs. It naturally accounts for the observations of [2] that show that the space density of BL Lacertae objects increases with cosmic time; in the opposite sense of the evolution of other Active Galactic Nuclei. Constrained by the luminosity function of QSOs at z= 2 and the requirement that their remnants at z < 1 be faint, we find that a power-law luminosity function that is relativistically beamed can give redshift and magnitude counts compatible with the observations. Our successful model has free parameters that were adjusted to yield a good fit to the observations, and the solution obtained may not be unique. On the other hand, the values used for these free parameters are all reasonable and compatible with the observations.

To obtain agreement between the observed space densities, we make the additional assumption that all QSOs, radio-quiet as well as radio-loud, are capable of generating jets, but that jets are snuffed in the young RQQs and only emerge in old age, thus implicitly assuming a unification of RLQs and RQQs.

The meager existing statistics of the classification of the host galaxies of BLLs seem to indicate that they favor elliptical (60%) rather than disk galaxies (30%). Prima facie, one would have to conclude that this is only allowed by our hypothesis either if elliptical can form RQQs as well as RLQs at z ~ 2 or if a large fraction of QSO-hosting spirals can evolve into elliptical. Recent HST observations seem to favor the later.

Although this paper does not prove that BLLs are quasar remnants, it shows that the hypothesis is compatible with our present knowledge of the statistics of QSOs and BL Lac objects, its outstanding success being its natural prediction that the space density of BLLs should increase with cosmic time, and that it should be taken seriously. It does not say anything regarding the unbeamed component of BL Lac objects, except that they should be optically inconspicuous objects and have FR I morphology.

**Aknowledgments**

This research was supported by the Natural Sciences and Engineering Research council of Canada and the FCAR program of the province of Québec. I wish to thank Dr. I. McHardy for sending me an unpublished HST image of PKS1413+135.

**FIGURE CAPTIONS**

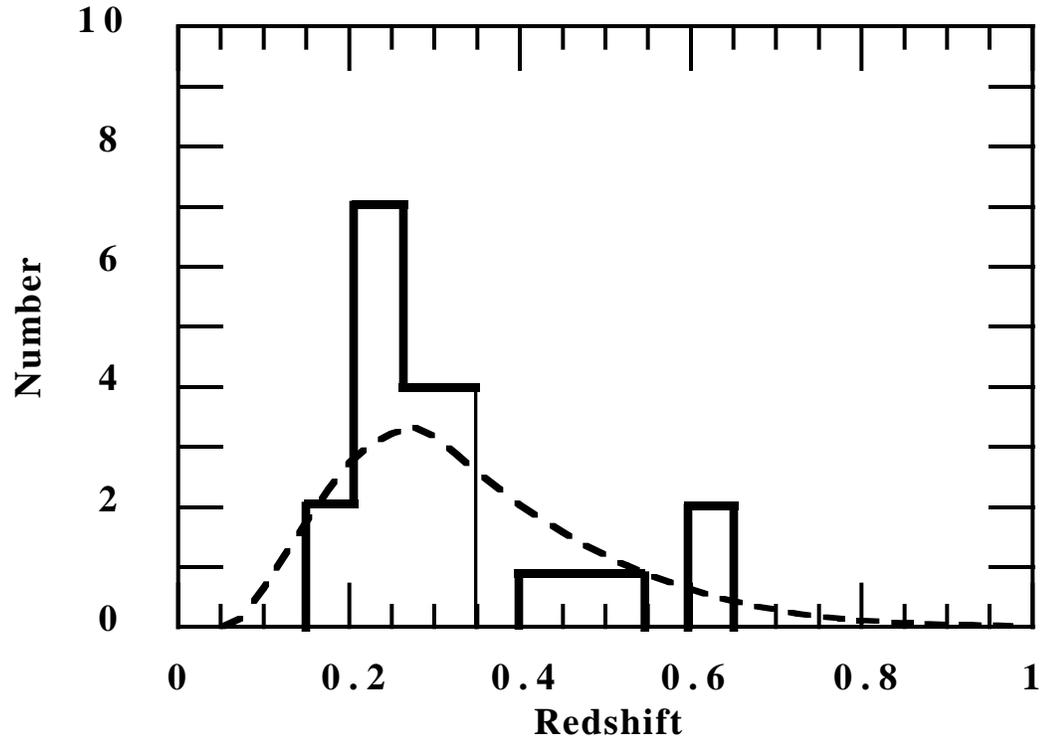

Figure 1: The dashed line gives the redshift distribution obtained from the relativistically beamed model described in the text. The counts are in bins of $\Delta z = 0.05$. The histogram shows the observations of [2].

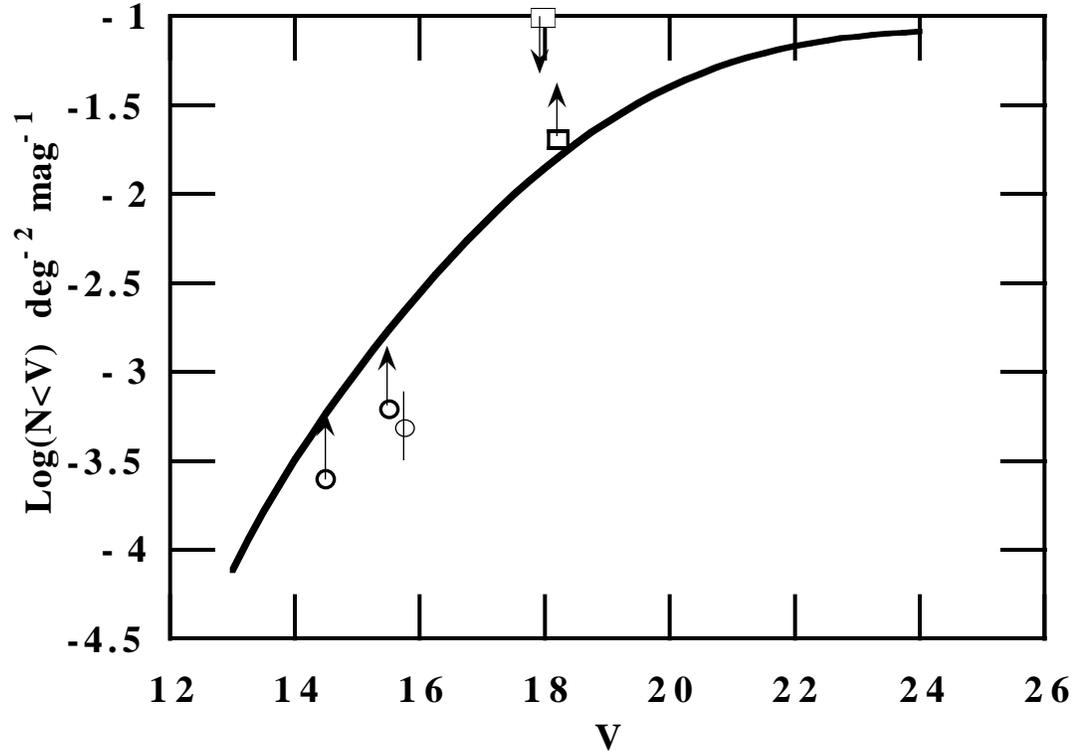

Figure 2: Integrated counts (per square degree) from the relativistically beamed model described in the text. The observed counts are from [6]; and the symbols have the same meaning as in that reference. The arrows indicate upper or lower limits.